\documentclass{webofc}
\usepackage[varg]{txfonts}   % Web of Conferences font
\begin{document}
\title{NIKA2 mapping and cross-instrument SED extraction of extended sources with Scanamorphos}

\author{\firstname{H.} \lastname{Roussel} \inst{\ref{IAP}}\fnsep\thanks{\email{roussel@iap.fr}}
\and \firstname{N.} \lastname{Ponthieu} \inst{\ref{IPAG}}
\and \firstname{R.} \lastname{Adam} \inst{\ref{LLR},\ref{CEFCA}}
\and \firstname{P.} \lastname{Ade} \inst{\ref{Cardiff}}
\and \firstname{P.} \lastname{Andr\'e} \inst{\ref{CEA1}}
\and \firstname{A.} \lastname{Andrianasolo} \inst{\ref{IPAG}}
\and \firstname{H.} \lastname{Aussel} \inst{\ref{CEA1}}
\and \firstname{A.} \lastname{Beelen} \inst{\ref{IAS}}
\and \firstname{A.} \lastname{Beno\^it} \inst{\ref{Neel}}
\and \firstname{A.} \lastname{Bideaud} \inst{\ref{Neel}}
\and \firstname{O.} \lastname{Bourrion} \inst{\ref{LPSC}}
\and \firstname{M.} \lastname{Calvo} \inst{\ref{Neel}}
\and \firstname{A.} \lastname{Catalano} \inst{\ref{LPSC}}
\and \firstname{B.} \lastname{Comis} \inst{\ref{LPSC}}
\and \firstname{M.} \lastname{De~Petris} \inst{\ref{Roma}}
\and \firstname{F.-X.} \lastname{D\'esert} \inst{\ref{IPAG}}
\and \firstname{S.} \lastname{Doyle} \inst{\ref{Cardiff}}
\and \firstname{E.~F.~C.} \lastname{Driessen} \inst{\ref{IRAMF}}
\and \firstname{A.} \lastname{Gomez} \inst{\ref{CAB}}
\and \firstname{J.} \lastname{Goupy} \inst{\ref{Neel}}
\and \firstname{F.} \lastname{K\'eruzor\'e} \inst{\ref{LPSC}}
\and \firstname{C.} \lastname{Kramer} \inst{\ref{IRAME}}
\and \firstname{B.} \lastname{Ladjelate} \inst{\ref{IRAME}}
\and \firstname{G.} \lastname{Lagache} \inst{\ref{LAM}}
\and \firstname{S.} \lastname{Leclercq} \inst{\ref{IRAMF}}
\and \firstname{J.-F.} \lastname{Lestrade} \inst{\ref{LERMA}}
\and \firstname{J.F.} \lastname{Mac\'ias-P\'erez} \inst{\ref{LPSC}}
\and \firstname{P.} \lastname{Mauskopf} \inst{\ref{Cardiff},\ref{Arizona}}
\and \firstname{F.} \lastname{Mayet} \inst{\ref{LPSC}}
\and \firstname{A.} \lastname{Monfardini} \inst{\ref{Neel}}
\and \firstname{L.} \lastname{Perotto} \inst{\ref{LPSC}}
\and \firstname{G.} \lastname{Pisano} \inst{\ref{Cardiff}}
\and \firstname{V.} \lastname{Rev\'eret} \inst{\ref{CEA1}}
\and \firstname{A.} \lastname{Ritacco} \inst{\ref{IRAME}}
\and \firstname{C.} \lastname{Romero} \inst{\ref{IRAMF}}
\and \firstname{F.} \lastname{Ruppin} \inst{\ref{MIT}}
\and \firstname{K.} \lastname{Schuster} \inst{\ref{IRAMF}}
\and \firstname{S.} \lastname{Shu} \inst{\ref{IRAMF}}
\and \firstname{A.} \lastname{Sievers} \inst{\ref{IRAME}}
\and \firstname{C.} \lastname{Tucker} \inst{\ref{Cardiff}}
\and \firstname{R.} \lastname{Zylka} \inst{\ref{IRAMF}}
}

\institute{\label{IAP} {\scriptsize Institut d'Astrophysique de Paris, Sorbonne Université, CNRS (UMR7095), 75014 Paris, France}
\vspace*{-0.1cm}
\and \label{IPAG} {\scriptsize Univ. Grenoble Alpes, CNRS, IPAG, 38000 Grenoble, France}
\vspace*{-0.1cm}
\and \label{LLR} {\scriptsize LLR (Laboratoire Leprince-Ringuet), CNRS, \'Ecole Polytechnique, Institut Polytechnique de Paris, Palaiseau, France}
\vspace*{-0.1cm}
\and \label{CEFCA} {\scriptsize Centro de Estudios de F\'isica del Cosmos de Arag\'on (CEFCA), Plaza San Juan, 1, planta 2, E-44001, Teruel, Spain}
\vspace*{-0.1cm}
\and \label{Cardiff} {\scriptsize Astronomy Instrumentation Group, University of Cardiff, UK}       
\vspace*{-0.1cm}
\and \label{CEA1} {\scriptsize AIM, CEA, CNRS, Universit\'e Paris-Saclay, Universit\'e Paris Diderot, Sorbonne Paris Cit\'e, 91191 Gif-sur-Yvette, France}
\vspace*{-0.1cm}
\and \label{IAS} {\scriptsize Institut d'Astrophysique Spatiale (IAS), CNRS and Universit\'e Paris Sud, Orsay, France}  
\vspace*{-0.1cm}
\and \label{Neel} {\scriptsize Institut N\'eel, CNRS and Universit\'e Grenoble Alpes, France}
\vspace*{-0.1cm}
\and \label{LPSC} {\scriptsize Univ. Grenoble Alpes, CNRS, Grenoble INP, LPSC-IN2P3, 53, avenue des Martyrs, 38000 Grenoble, France}
\vspace*{-0.1cm}
\and \label{Roma} {\scriptsize Dipartimento di Fisica, Sapienza Universit\`a di Roma, Piazzale Aldo Moro 5, I-00185 Roma, Italy}
\vspace*{-0.1cm}
\and \label{IRAMF} {\scriptsize Institut de RadioAstronomie Millim\'etrique (IRAM), Grenoble, France}
\vspace*{-0.1cm}
\and \label{CAB} {\scriptsize Centro de Astrobiolog\'ia (CSIC-INTA), Torrej\'on de Ardoz, 28850 Madrid, Spain}
\vspace*{-0.1cm}
\and \label{IRAME} {\scriptsize Instituto de Radioastronom\'ia Milim\'etrica (IRAM), Granada, Spain}
\vspace*{-0.1cm}
\and \label{LAM} {\scriptsize Aix Marseille Univ, CNRS, CNES, LAM (Laboratoire d'Astrophysique de Marseille), Marseille, France}
\vspace*{-0.1cm}
\and \label{LERMA} {\scriptsize LERMA, Observatoire de Paris, PSL Research University, CNRS, Sorbonne Universit\'e, 75014 Paris, France}
\vspace*{-0.1cm}
\and \label{Arizona} {\scriptsize School of Earth and Space Exploration and Department of Physics, Arizona State University, Tempe, AZ 85287} 
\vspace*{-0.1cm}
\and \label{MIT} {\scriptsize Kavli Institute for Astrophysics and Space Research, Massachusetts Institute of Technology, Cambridge, MA 02139, USA \\ ~}
}

\abstract{%
The steps taken to tailor to NIKA2 observations the Scanamorphos algorithm (initially developed to subtract low-frequency noise from {\it Herschel} on-the-fly observations) are described, focussing on the consequences of the different instrument architecture and observation strategy. The method, making the most extensive use of the redundancy built in the multi-scan coverage with large arrays of a given region of the sky, is applicable to extended sources, while the pipeline is so far optimized for compact sources. An example of application is given.
A related tool to build consistent broadband SEDs from 60\,microns to 2\,mm, combining {\it Herschel} and NIKA2 data, has also been developed. Its main task is to process the data least affected by low-frequency noise and coverage limitations (i.e. the {\it Herschel} data) through the same transfer function as the NIKA2 data, simulating the same scan geometry and applying the same noise and atmospheric signal as extracted from the 1\,mm and 2\,mm data.
}
\maketitle
%
%For bibliography use \cite{RefJ}

\section{Introduction}
\label{intro}
%%%
Infrared and millimeter arrays made of hundreds to thousands of detectors, such as NIKA2, allow very efficient on-the-fly mapping of large fields of view, but call for special data processing techniques to remove both the atmospheric foreground and the instrumental low-frequency noise, hereafter generically called drifts, or low-frequency noise, while preserving extended astrophysical sources. We here present a tool designed to extract all the drift components from the observations by exploiting the redundancy as thoroughly as possible, taking advantage of the fact that within the nominal field of view, each position on the sky is sampled by many different detectors at many different times, and that the atmosphere varies significantly between successive scans of the same source. The underlying assumptions are the same as for the version of the tool initially developed for PACS and SPIRE onboard {\it Herschel}, despite the obvious differences in the physical nature of the drifts and in the instrument architecture: (1) the astrophysical signal is invariant while the drifts are supposed to vary on all timescales comprised between about 10 sampling periods and the map width crossing time~; (2) the probability distribution of the uncorrelated part of the drifts is symmetric about zero, but not necessarily Gaussian.

The effort to adapt the algorithm to NIKA2 observations is still underway, and for the moment applies to total-intensity observations only. It is part of a larger endeavor to apply it to various ground-based and balloon-borne instruments, all affected by dominant low-frequency noise: the ArT\'eMiS 350-450\,$\mu$m camera installed at APEX \cite{Andre16} and the PILOT experiment in polarization at 240\,$\mu$m, operating at altitudes of $\simeq 35 - 40$\,km \cite{Mangilli19}.

%%%
%%%
\section{Principles and major steps}
\label{descr}

\subsection{Prerequisites}
\label{pre}
A thorough description of the instrument, the observing modes and the whole calibration procedure is given by \cite{Adam18} and \cite{Perotto19}. The raw data have to be pre-processed with the pipeline to accomplish these tasks: detector identification, opacity correction, flux calibration, computation of spatial coordinates, and rejection of the few detectors with overlapping frequency tones. Then they are reformatted, with basic information attached, by dedicated scripts.

\subsection{Relative gains for extended emission}
\label{gains}
All multiplicative effects must be corrected before the low-frequency noise can be subtracted, since an incorrect calibration will induce errors in the signal of the same source seen by different detectors, that will be ascribed to drifts and therefore compromise flux preservation. The flux calibration is done for point sources from special products called beammaps, allowing to obtain a well-sampled map of a primary calibrator for each individual detector \cite{Perotto19}, but this procedure does not allow to capture variations in the ratio between the extended part and the core of the beam from detector to detector. Hence, to calibrate the relative responses to extended emission, an additional step is necessary. A simple way of doing it is to use the responses to the atmosphere, since it is very bright, can be considered uniform across the array size (which was verified), and is available in all on-the-fly observations, allowing to check the stability of the responses during a whole campaign. In practice, we use all scans longer than 5 min to iteratively estimate the atmospheric signal and fit affine functions of this signal for each valid detector. Since the gains are much better constrained when the atmosphere fluctuates a lot with time, the global gains of the campaign are computed by weighing the gains of each scan by the dynamic range of the atmosphere brightness. At 1mm, the atmospheric signal is determined from array 3, that is much better behaved than array 1. Consequently, the global relative response of array 1 is not unity (unlike for arrays 2 and 3, by construction), but on the order of 1.15\,.
Although in principle these near-field responses are not exactly equivalent to the far-field responses that are truly needed, it is better to take them into account: their application improves the results obtained on bright extended sources at 1mm.

\subsection{Algorithm and step-by-step processing}
\label{steps}
The drifts can be decomposed according to their physical origin: the atmospheric emission, strongly correlated between all detectors~; the electronic noise, strongly correlated between detectors belonging to the same electronic box or the same sub-box \cite{Ponthieu19}~; and the uncorrelated flicker noise. However, to avoid having to determine the correlation coefficients for all detectors, it is much simpler from an algorithmic point of view to adopt a different decomposition: the average drift for the whole array~; the average drift for each electronic box or sub-box~; and the complement, i.e. the individual drifts. No relationship is assumed between the atmospheric emission at 1mm and 2mm, and the processing in one band is totally independent from the other band. Nevertheless, the average drift (heavily dominated by the atmosphere) extracted from the data is consistent with the same function in both bands (with normalizations that are constant under similar observing conditions).

The corrections using the redundancy (i.e. numbers 2 to 4 below) make use of a coarse spatial grid, where the size of a coarse pixel is equal to the stability length (typically between 0.5 and 1 times the FWHM), corresponding to the minimum timescale $t_c$ below which the drifts will not be corrected. This stability length must be large enough to allow computing simple statistics from the samples of each detector taken during such a time interval. The requirement is that the minimum number of samples per crossing is 7. These statistics allow rejecting data having fluctuations in significant excess of the white noise, either because of glitches or because of contamination by a compact source. For more details than is practical to give here, the reader is referred to \cite{Roussel13}.

The recorded signal is modeled as:~~ $R(t, k_i) = S(p) + D_{\rm aver}(t) + D_{\rm indiv}(t, k_i) + HF(t, k_i)$ \\
as a function of time $t$, detectors $k_i$ and coarse pixels $p$, where $S$ is the sky signal, $HF$ is the high-frequency noise (white noise and glitches), and $D$ designates the drifts.

Since the noise has more power at the longest temporal and spatial scales, for each noise component the longest timescales have to be subtracted first. Here are the main steps, in the order in which they are performed: \\
{\bf (1)} Baseline subtraction, on timescales comprised between the map crossing time and the scan duration: Baselines consist of linear fits of the signal averaged over the whole array or over electronic boxes, as a function of time, and are designed to be robust with respect to outlying samples. The correction is iterative, and a source mask automatically built from the current map is updated at each iteration and transferred to the time domain. The criteria for the mask are based on thresholds in units of the error map and of the standard deviation in the map, as well as the need to avoid masking too large a fraction of the map or masking predominantly the edges. \\
{\bf (2)} Average drift subtraction on short timescales: Within each coarse pixel, differences between pairs of detector crossings are computed and coadded according to their times:
$\Delta(t_1, t_2) = R(t_1, k_i) - R(t_2, k_j) = D_{\rm aver}(t_1) - D_{\rm aver}(t_2) + D_{\rm indiv}(t_1, k_i) - D_{\rm indiv}(t_2 , k_j)$.
After coaddition over all coarse pixels and detectors, given enough redundancy, the differences of individual drifts vanish, and one obtains differences between the average drift at two different times.
These differences are stored in a matrix, that is scanned several times to iteratively build the time series of the average drift, assuming a null median. Since the solution is not unique, there is a spurious excess drift, of the same periodicity as the geometry of the observations, that is removed by another instance of the baseline subtraction, optionally including piecewise baselines for very large fields (but the same function for each leg of a scan). \\
{\bf (3)} Average drift per electronic box or sub-box, on successively smaller timescales: This component is subtracted after computing the individual drifts (as in step 4), and then their weighted average for each box or sub-box separately. \\
{\bf (4)} Individual drifts, on successively smaller timescales: Within each coarse pixel, they are computed as the difference between each detector crossing and the weighted average of all crossings:~~
$\delta(t, k_i) = R(t, k_i) - 1/N \sum_{j=1,...N} R(t_j, k_j) = D_{\rm aver}(t) + D_{\rm indiv}(t, k_i) - 1/N \sum_{j=1,...N} (D_{\rm aver}(t_j) + D_{\rm indiv}(t_j, k_j))$~~
which reduces to $D_{\rm indiv}(t, k_i)$ provided there is enough redundancy, given that the average drift has been subtracted beforehand.
For each timescale longer than $t_c$ (corresponding to the stability length), the correction is binned in time, and steps 3 and 4 are performed successively and followed by a new instance of the baseline subtraction. When working at timescale $t_c$, only step 4 is done.

Other functionalities include the detection and masking of bad detectors, glitches, and instabilities (detuning) that are short enough to be treated as glitches, such as those frequently occurring during the March 2016 campaign, and the detection of scans where a large fraction of the detectors are not properly tuned for a large fraction of the time.

\begin{figure*}
\vspace*{-1.2cm}
\centering
\includegraphics[scale=0.15]{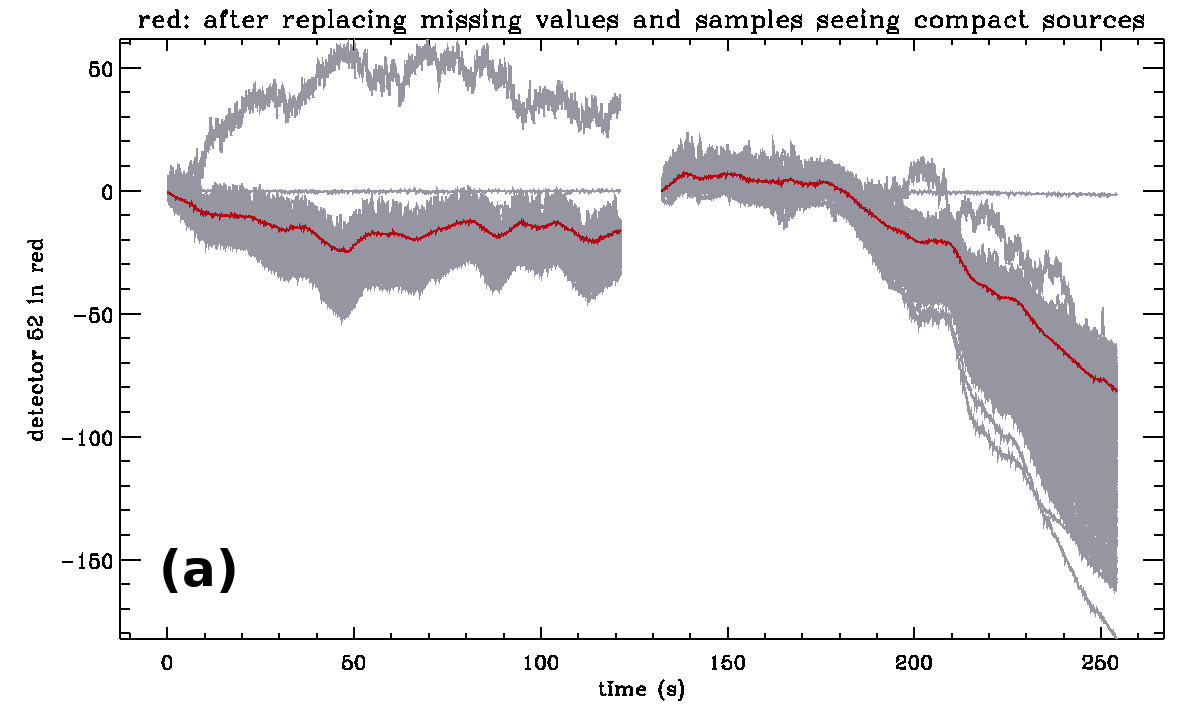}
\includegraphics[scale=0.15]{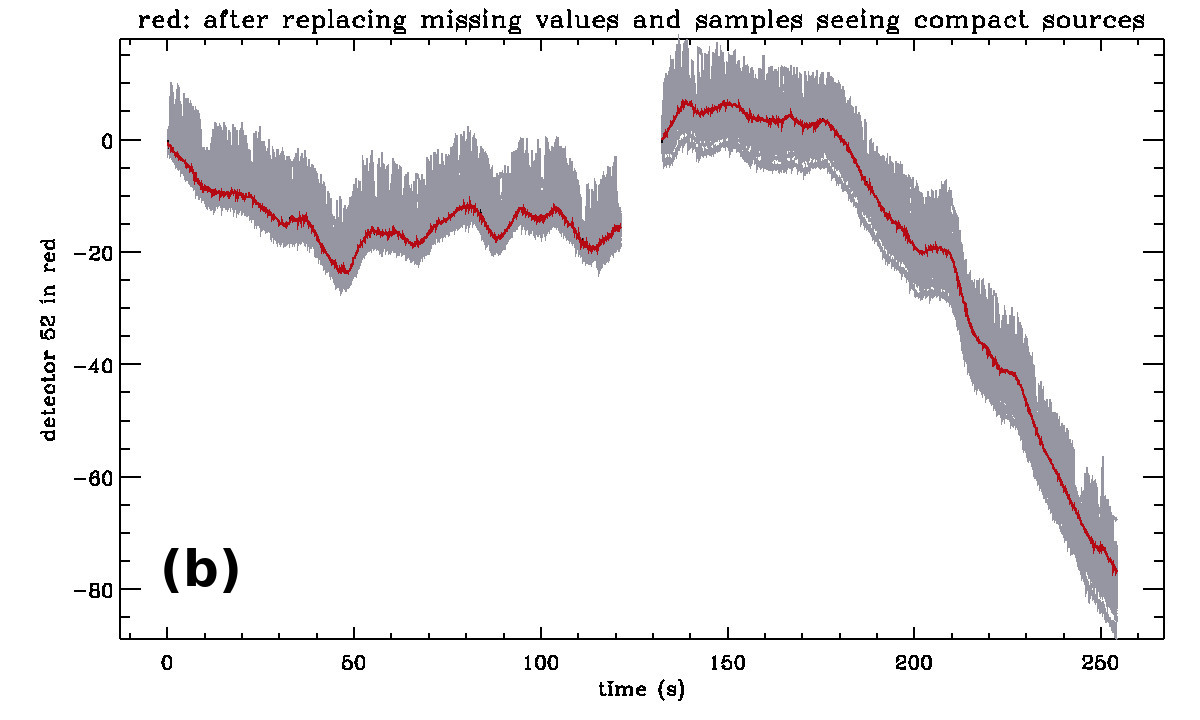}
\includegraphics[scale=0.15]{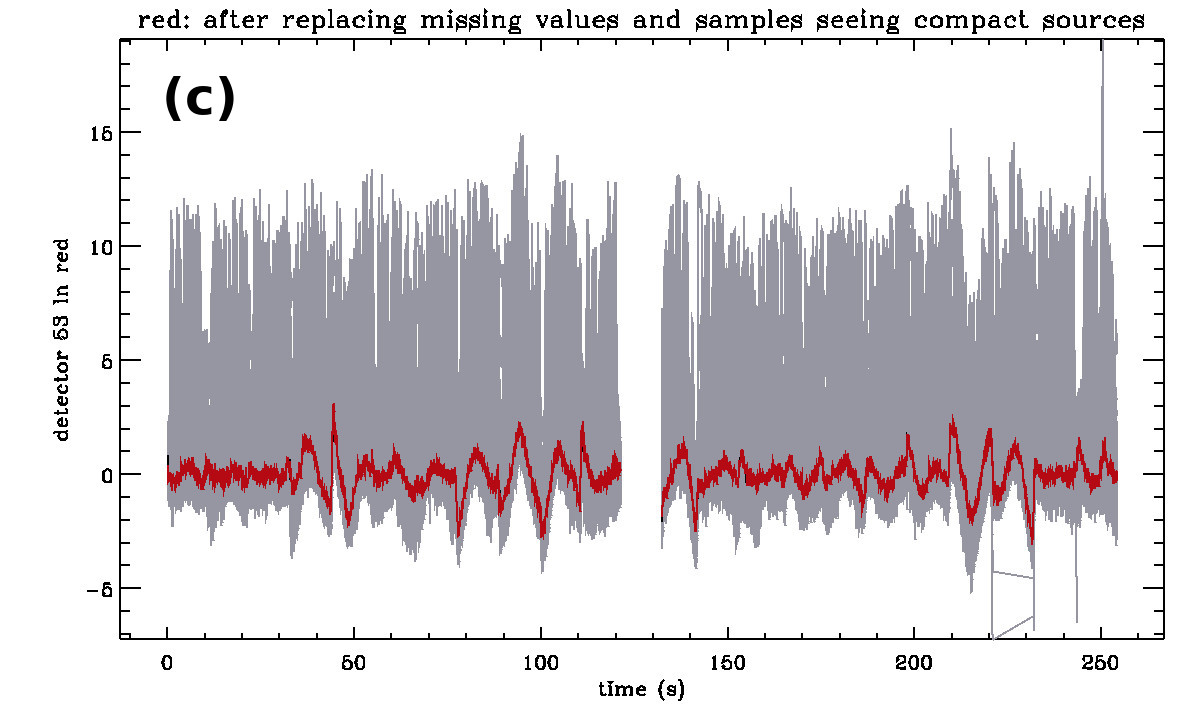}
\includegraphics[scale=0.15]{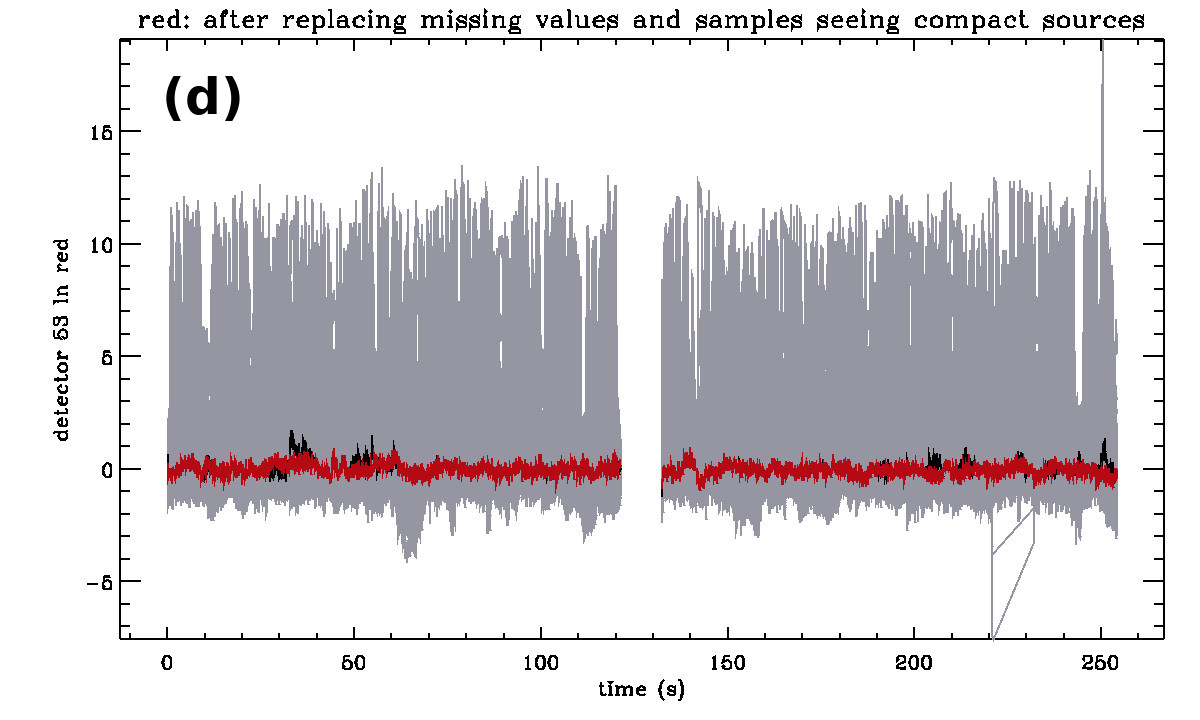}
\vspace*{-0.3cm}
\caption{Time series of all detectors in grey, and of a particular detector in red at various steps: (a) initially~; (b) after correction for the relative gains and masking of bad detectors~; (c) after first baseline subtraction~; (d) after correction for the average drift on small timescales. In these plots, samples where a compact source is detected are replaced with simulated white noise.}
\label{fig_timeseries}
\end{figure*}

\begin{figure*}
\vspace*{-0.5cm}
\centering
\includegraphics[scale=0.15]{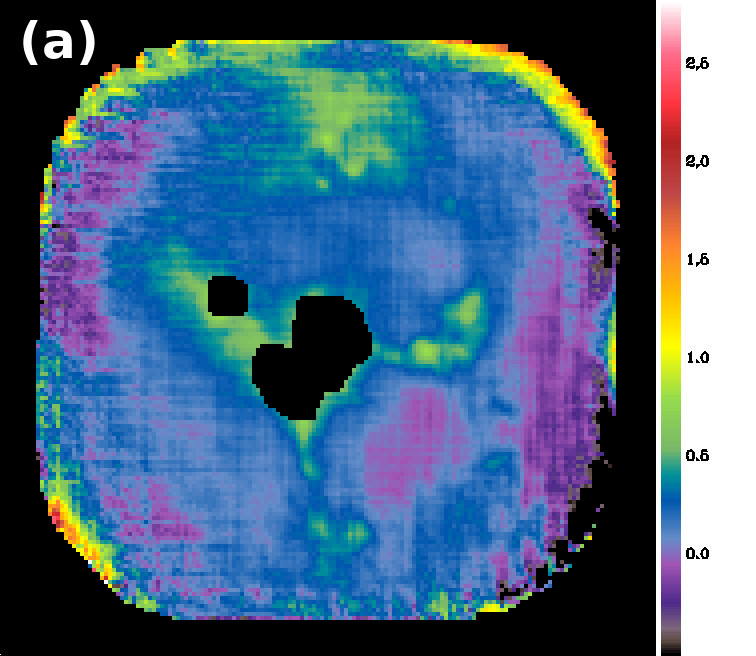}
\includegraphics[scale=0.15]{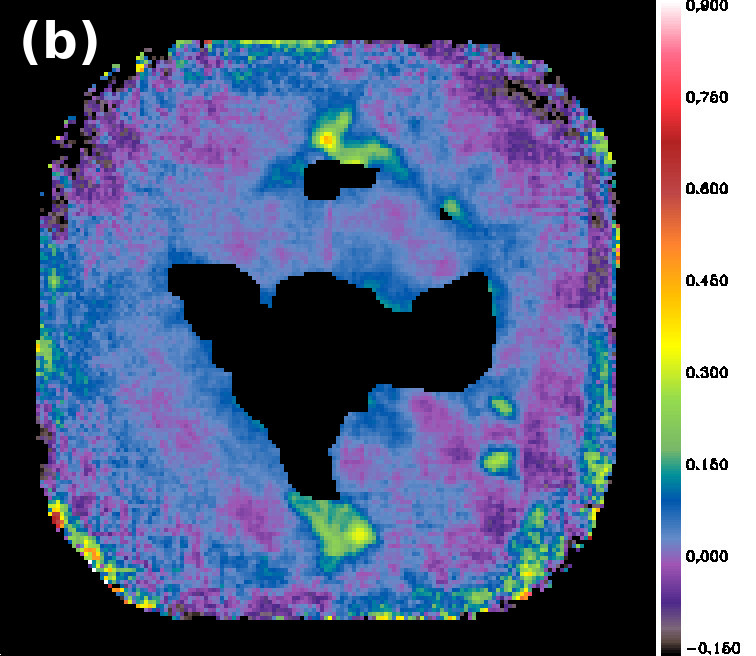}
\includegraphics[scale=0.15]{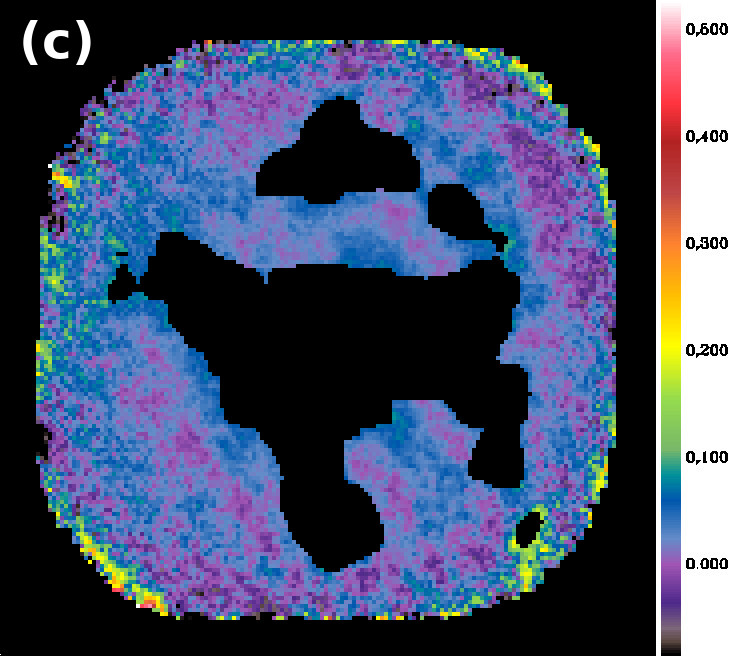}
\includegraphics[scale=0.15]{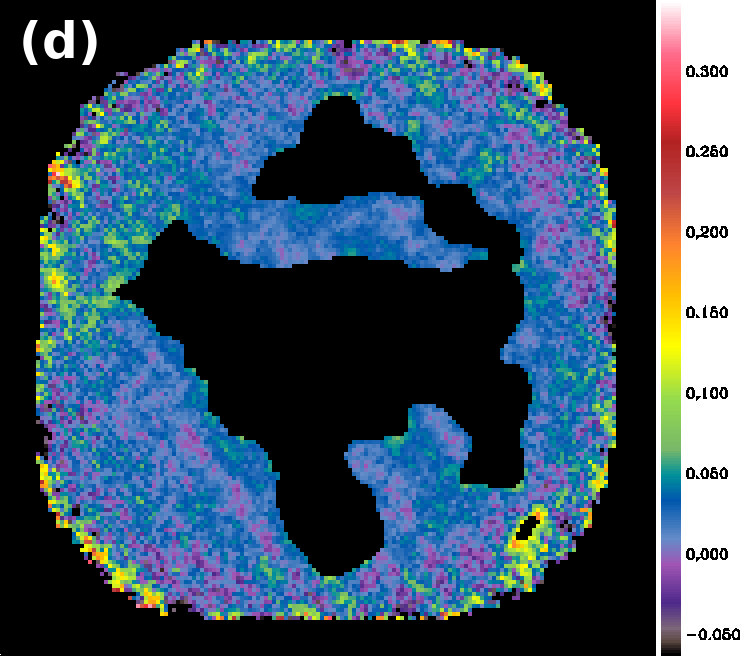}
\vspace*{-0.1cm}
\caption{Source mask automatically built during baseline subtraction at various steps: (a) after first baseline subtraction~; (b) after correction for the average drift on small timescales~; (c) after correction for the drifts on the $3.11$\,s timescale~; (d) after correction for the drifts on the $1.04$\,s timescale.}
\label{fig_sourcemask}
\vspace*{-0.5cm}
\end{figure*}

\begin{figure*}
\vspace*{-1cm}
\centering
\sidecaption
\includegraphics[scale=0.15]{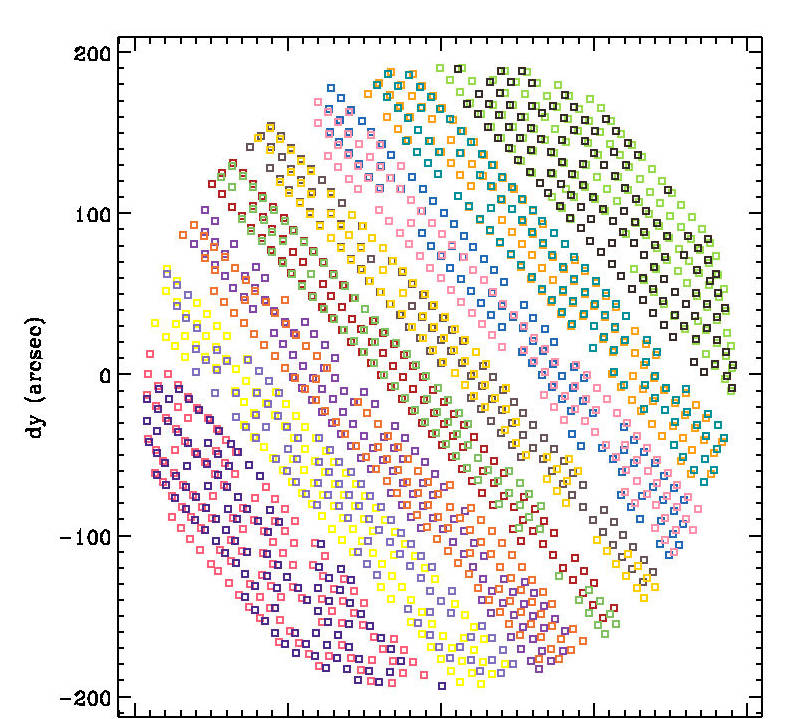}
\hspace*{0.3cm}
\includegraphics[scale=0.15]{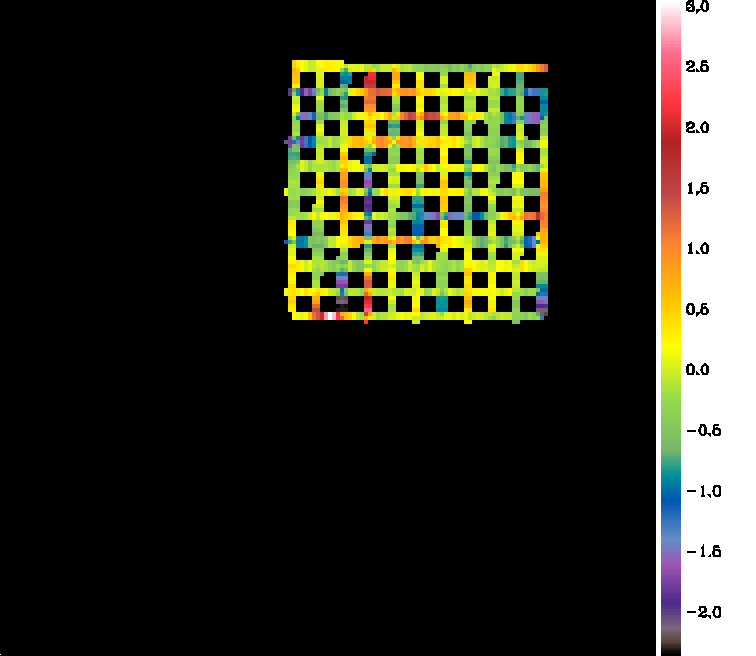}
\caption{Array geometry: (Left) The detectors of each electronic box of the two 1\,mm arrays are represented with a different color. (Right) Map of a single detector after baseline subtraction (including the two scans).}
\label{fig_geom}
\end{figure*}

\begin{figure*}[!ht]
\vspace*{-0.5cm}
\centering
\includegraphics[scale=0.15]{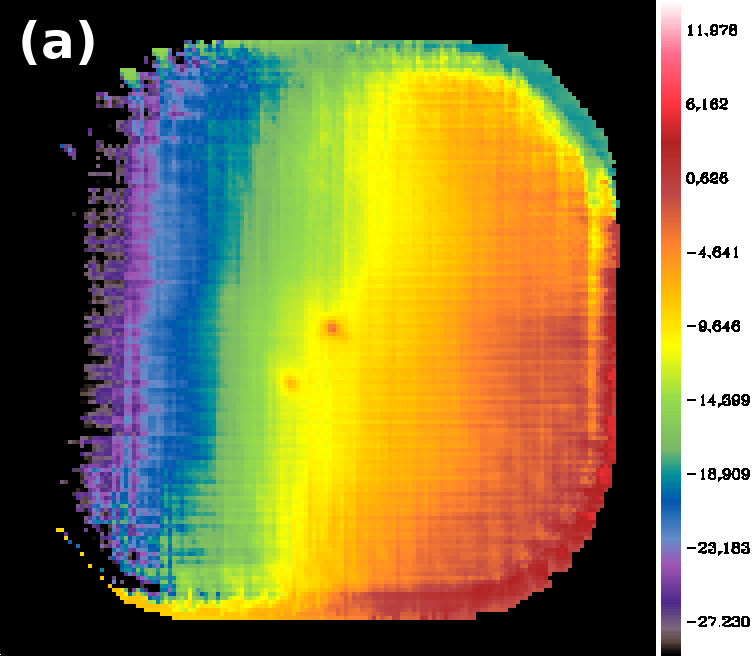}
\includegraphics[scale=0.15]{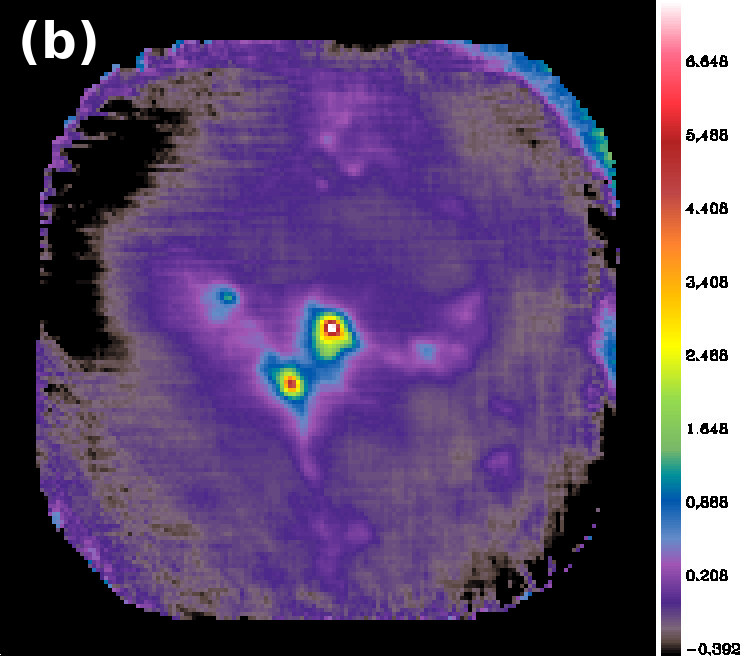}
\includegraphics[scale=0.15]{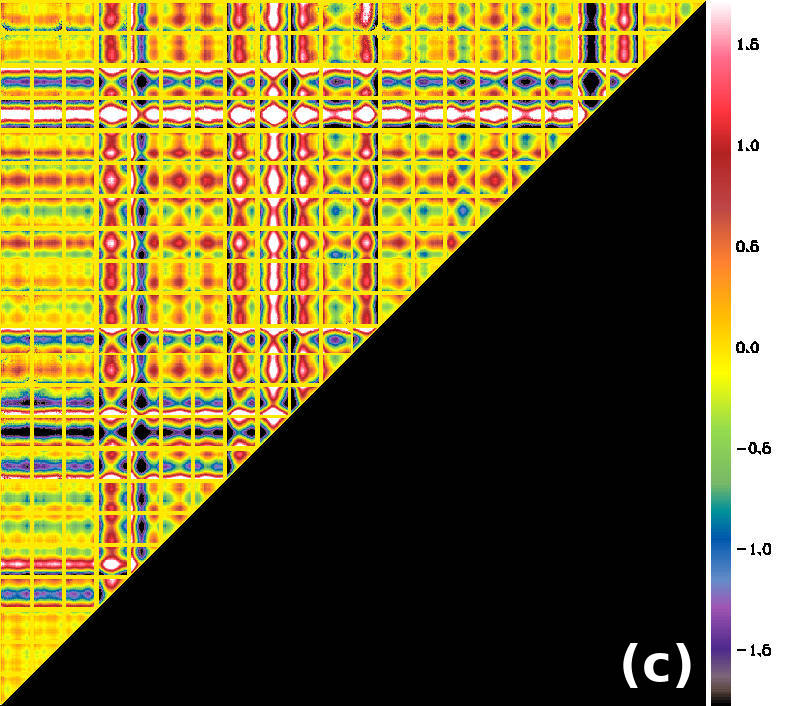}
\includegraphics[scale=0.15]{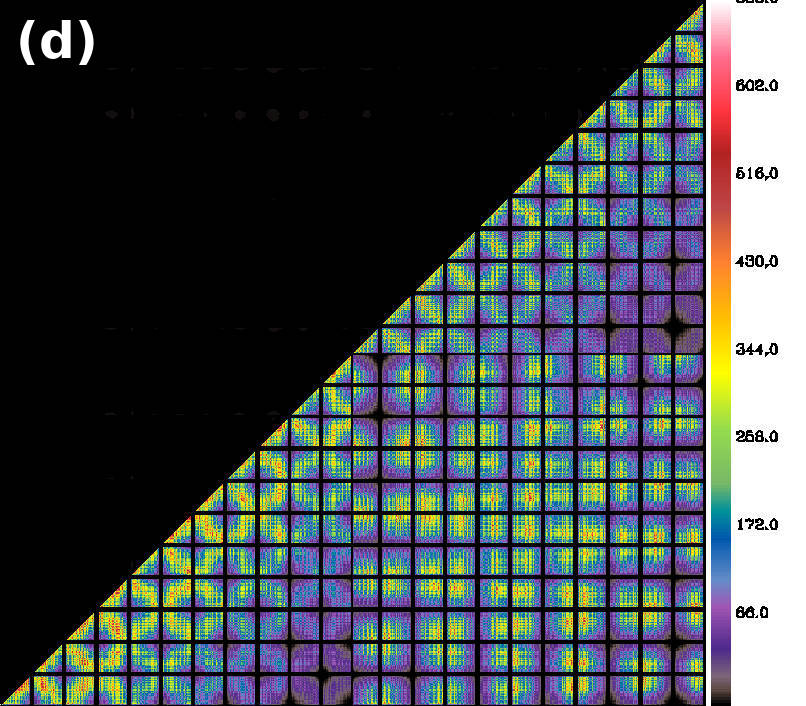}
\includegraphics[scale=0.15]{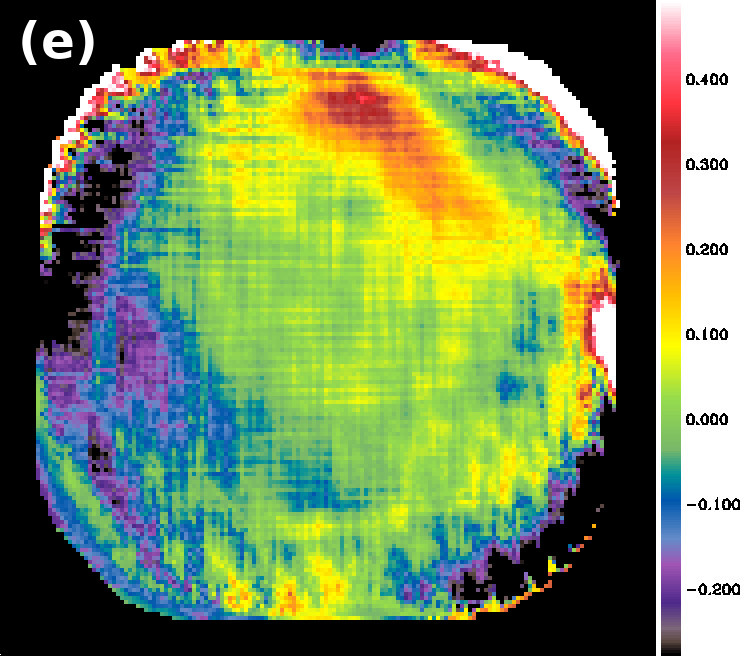}
\includegraphics[scale=0.15]{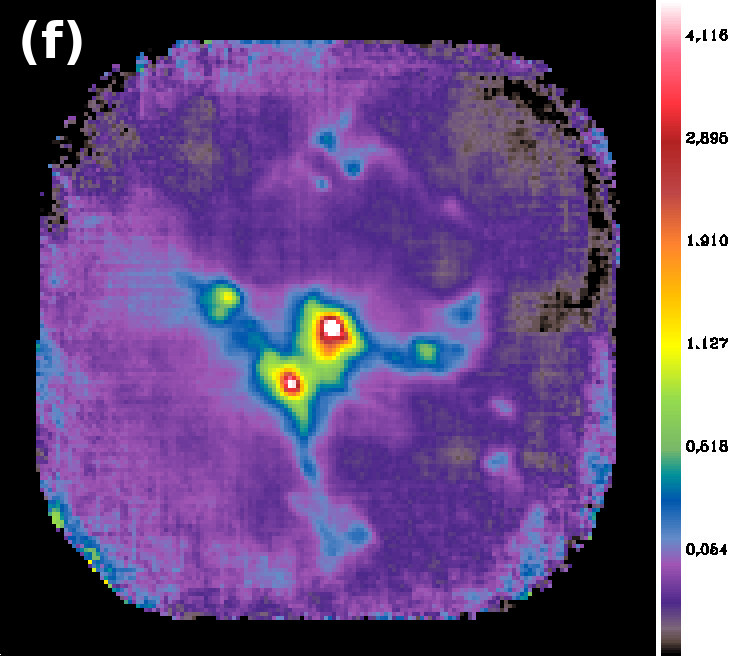}
\includegraphics[scale=0.15]{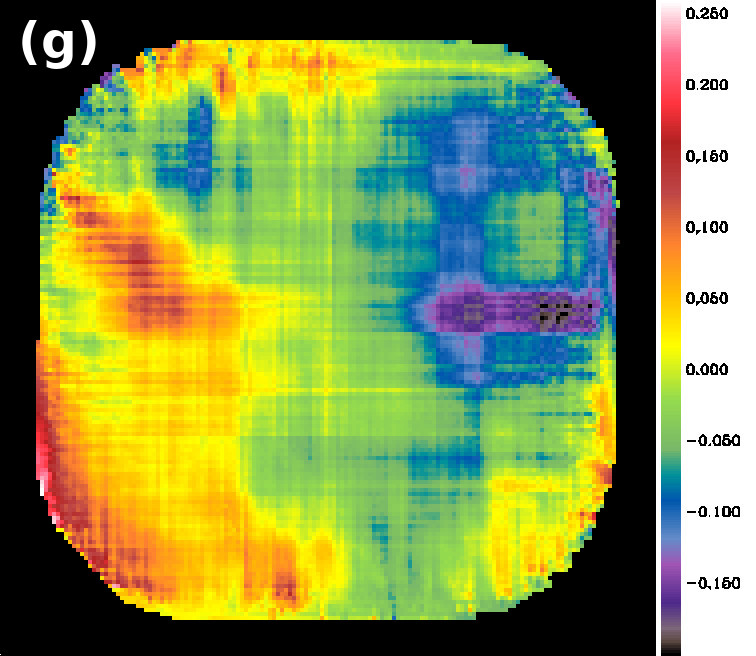}
\includegraphics[scale=0.15]{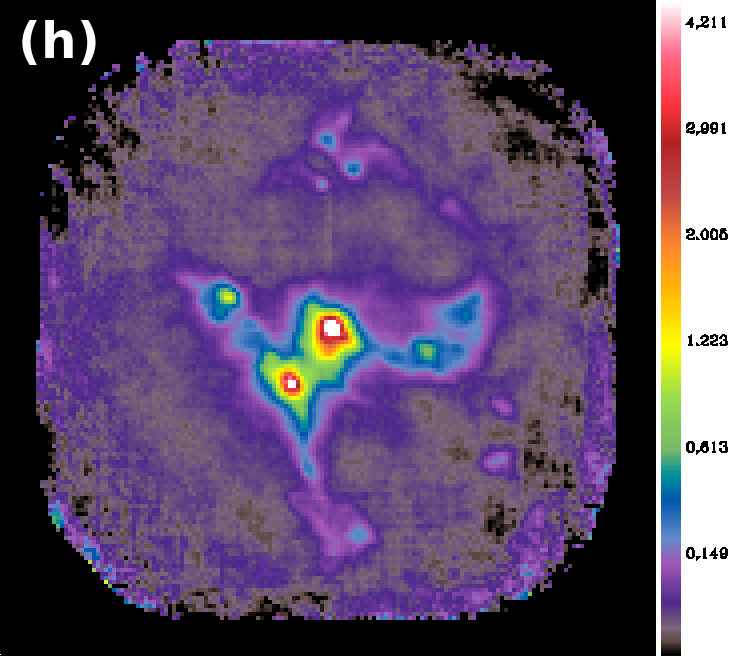}
\includegraphics[scale=0.15]{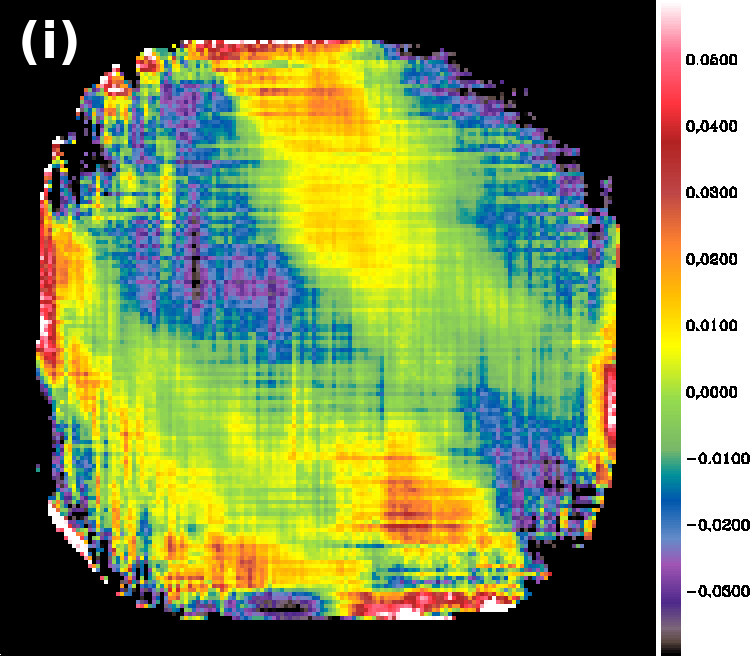}
\includegraphics[scale=0.15]{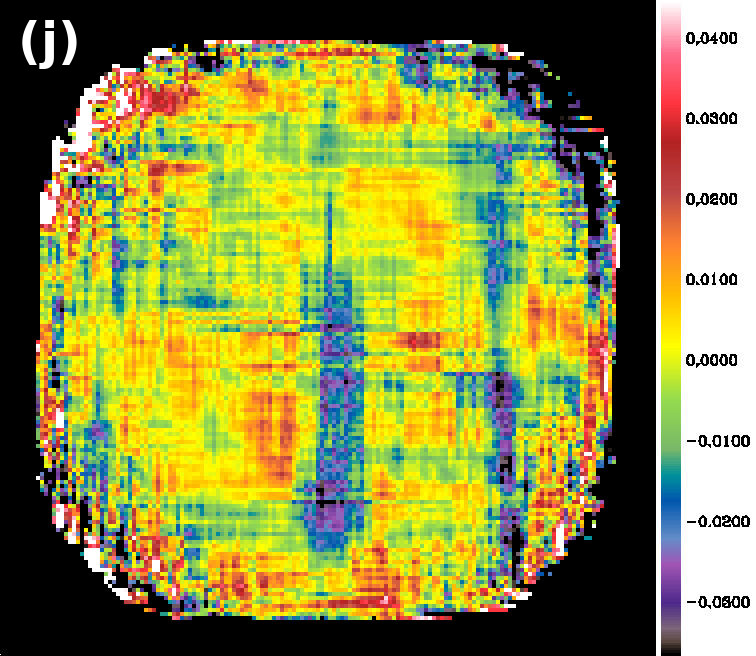}
\includegraphics[scale=0.15]{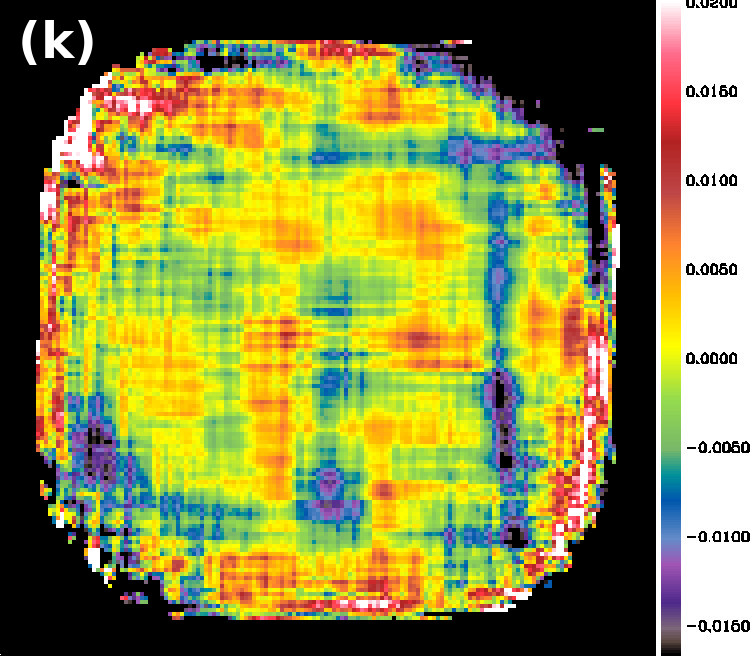}
\includegraphics[scale=0.15]{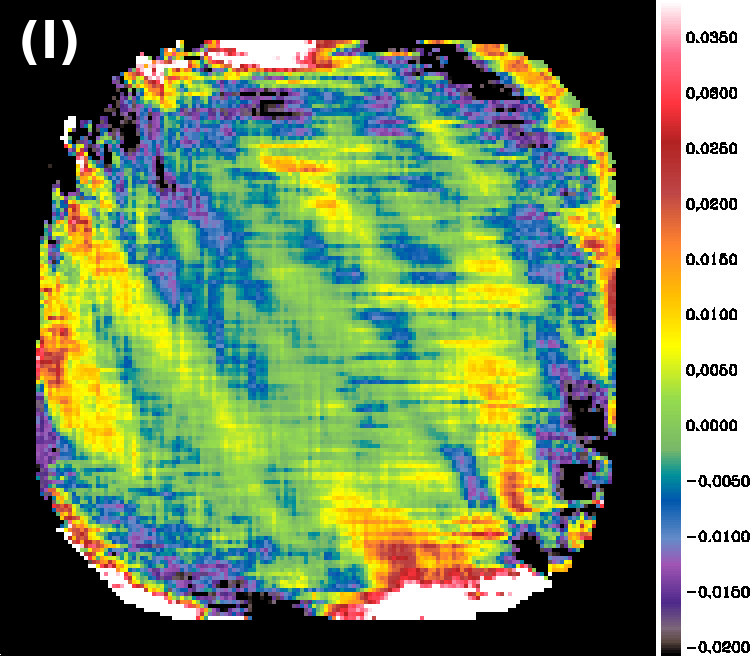}
\includegraphics[scale=0.15]{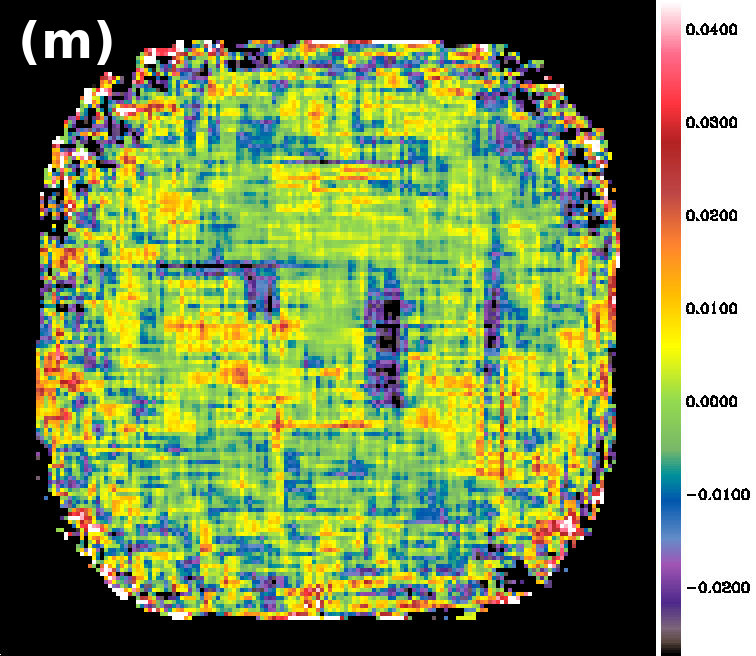}
\hspace*{0.5cm}
\includegraphics[scale=0.15]{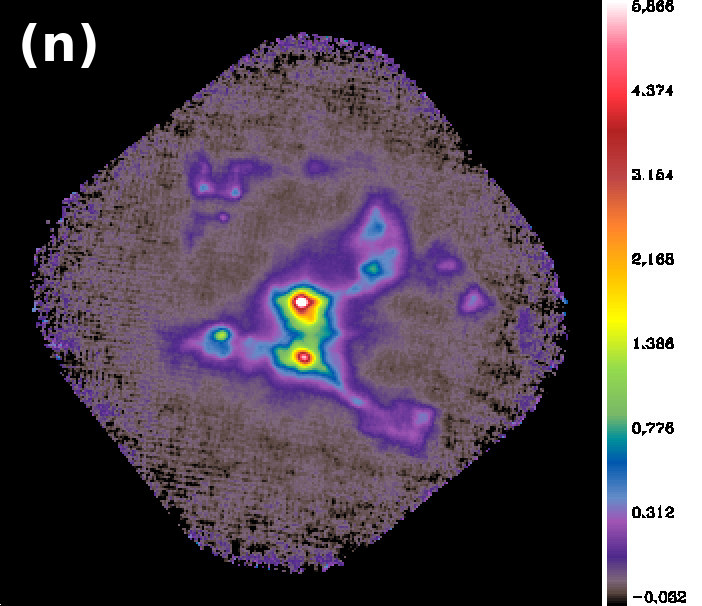}
\includegraphics[scale=0.15]{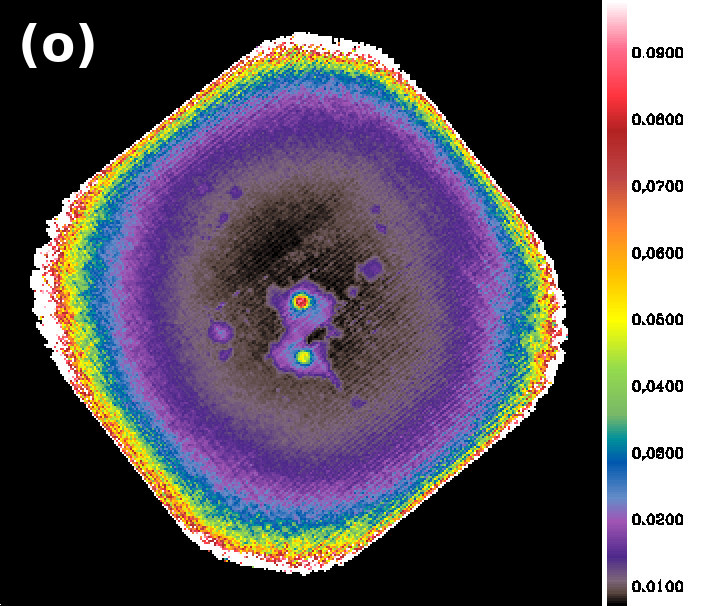}
\includegraphics[scale=0.15]{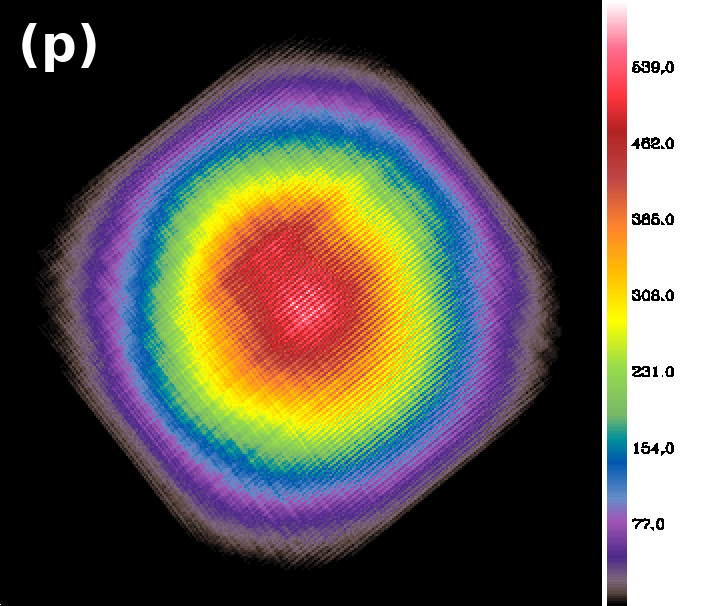}
\vspace*{-0.1cm}
\caption{Intermediate and final maps: (a) Raw data~; (b) After first baseline subtraction~; (c) Matrix of the average drift differences $\Delta(t_1, t_2)$~; (d) Weights associated with $\Delta(t_1, t_2)$~; (e) Projection of the subtracted average drift on small timescales~; (f) After subtraction of the average drift~; (g) Projection of the baselines computed after step f~; (h) After second baseline subtraction~; (i) Projection of the average drift per electronic box, on the $3.11$\,s timescale~; (j) Projection of the average drift per electronic sub-box, on the $3.11$\,s timescale~; (k) Projection of the individual drifts, on the $3.11$\,s timescale~; (l) Projection of the average drift per electronic box, on the $1.04$\,s timescale~; (m) Projection of the individual drifts, on the $1.04$\,s timescale~; (n) Final map~; (o) Error map~; (p) Weight map.}
\label{fig_steps}
\end{figure*}

\begin{figure*}
\vspace*{-1cm}
\centering
\hspace*{-5cm}
\includegraphics[scale=0.13]{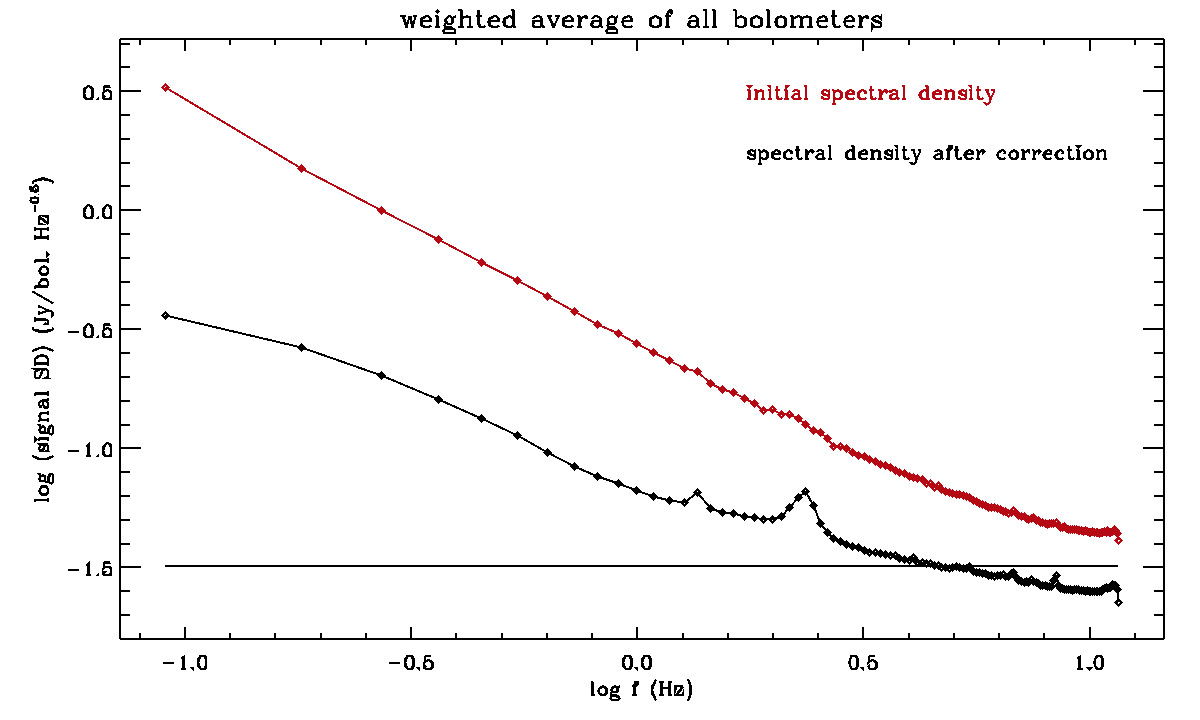}
\includegraphics[scale=0.13]{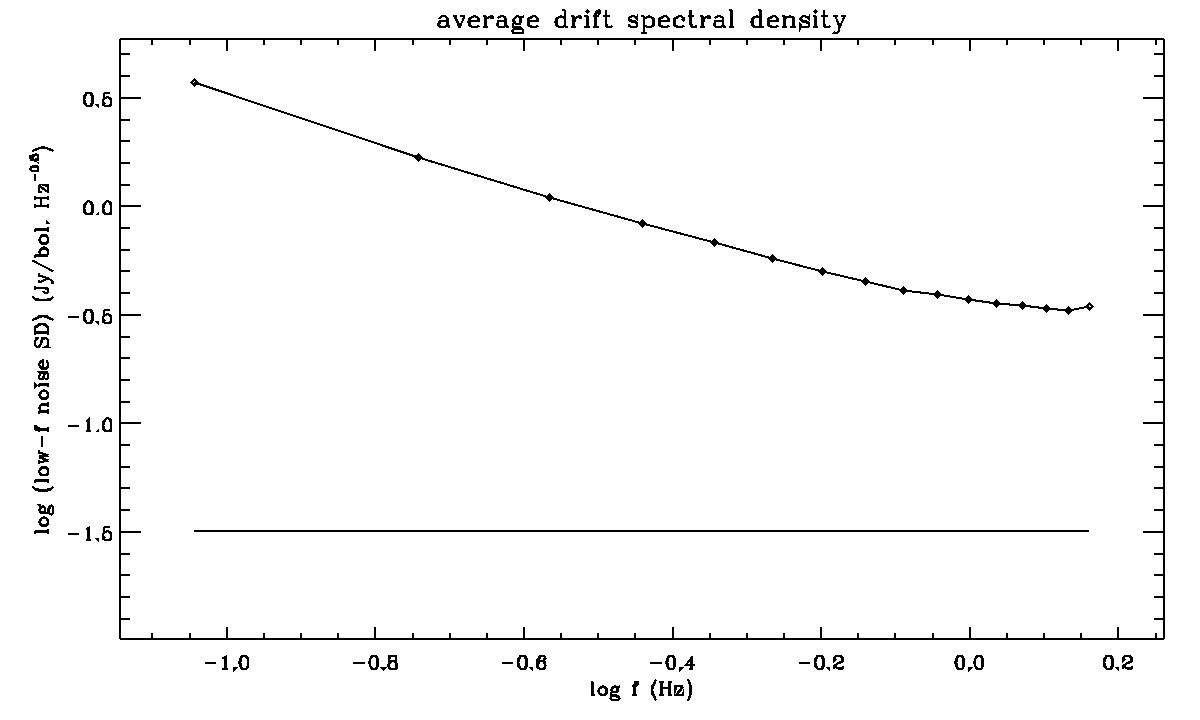}
\includegraphics[scale=0.13]{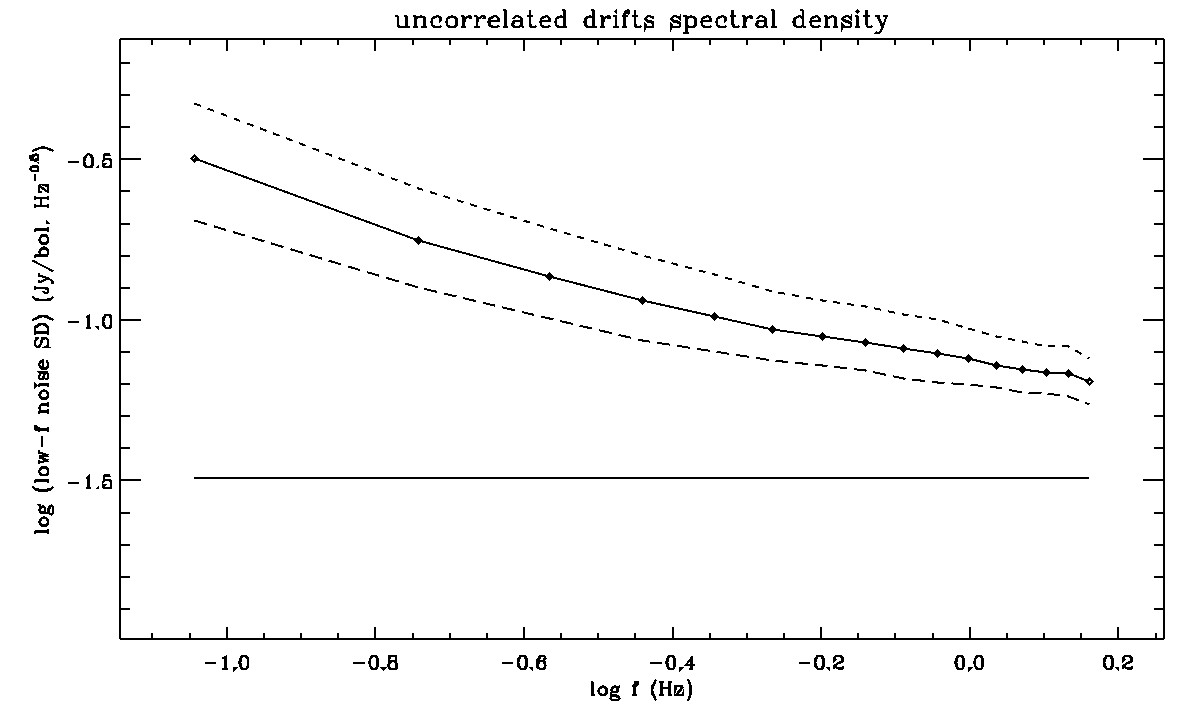}
\hspace*{-5cm}
\vspace*{-0.3cm}
\caption{Spectral densities of: (a) the average signal, before and after the processing~; (b) the average drift~; (c) the individual drifts.}
\label{fig_spd}
\vspace*{-0.5cm}
\end{figure*}

An illustration of the main steps at 1\,mm is shown in Fig.~\ref{fig_timeseries} to \ref{fig_spd}, for an extended bright source observed with only two scans at $30$\,arcsec\,s$^{-1}$, NGC\,7538\,. The separation between scan legs was relatively large, $30$\,arcsec, which might be the reason for the residual graininess of the sky. Note that the intermediate maps created during the processing are projected on a grid oriented along the scan directions. The median duration of a scan leg is $11.18$\,s, and the timescales for the subtraction of the individual drifts are $3.11$\,s, $1.04$\,s and $0.35$\,s.
\vspace*{-1cm} ~

\newpage

\vspace*{-0.8cm}

\subsection{Observing strategy requirements}

It is important to realize that the quality of the processing depends crucially on the way the observations are performed. The most delicate step, i.e. most sensitive to the observing strategy, is probably the baseline subtraction, and particularly the construction of the source mask. In order to optimize the separation between low-frequency noise and signal, the following guidelines, stemming from close inspection of commissioning data, should be applied: \\
{\bf (1)} The trajectories of the detectors have to vary as much as possible from one scan to the next, ensuring that the time series of both the astrophysical signal and the atmosphere do not repeat themselves. In practice, it is convenient to use two orthogonal scan directions (or, in case the source geometry prevents it, two directions making between them an angle greater than $\simeq 30$ degrees and as large as possible), and to keep alternating between them.
{\bf (2)} If the source to be mapped is elongated, it is best to avoid scanning along its major axis.
{\bf (3)} The scanning direction has to make a substantial angle with respect to the long axis of the electronic boxes (fixed in Nasmyth coordinates), to ensure a better determination of the average drift per electronic box, because in that case different positions on the sky are sampled at successive times for each box.
{\bf (4)} To ensure robust baselines, all detectors should go slightly beyond the source on each side of the scan, i.e. the scan length should be equal to the source size $+$ twice the array diameter along the scan direction.
{\bf (5)} To also allow sufficient coverage and redundancy in the region of interest, the map width should be significantly larger than the source size in the scan transverse direction. The width to be chosen is a compromise between coverage uniformity and observing time.
{\bf (6)} For low-surface brightness sources, it is useless to keep accumulating data when the effective opacity (at the elevation of the source) becomes greater than $\simeq 0.4$\,. For bright sources, effective opacities as high as $0.7$ can be tolerated.
{\bf (7)} During the observations, a vigilant eye should be kept on the tuning of the detectors. Samples taken out of resonance are in practice unusable.

%Don't forget to give each section, subsection, subsubsection, and
%paragraph a unique label (see Sect.~\ref{sec-1}).

\section{To conclude}
\label{conclusion}

Applications range from planets and secondary calibrators to star-forming molecular clouds and much fainter nearby galaxies (see the slides associated with these proceedings).

The machinery has been developed to simulate and process data from other instruments as if they had been acquired with the same geometry and the same noise as in NIKA2 data, i.e. applying the same transfer function. This will allow to combine {\it Herschel} and NIKA2 data of the same extended targets in order to obtain homogeneous spectral energy distributions. For lack of space, its description is postponed to another document.

The code will be made public after being vetted by the NIKA2 core team and will then be maintained and revised throughout the instrument operation. \\

{\scriptsize {\linespread{0.8}
We would like to thank the IRAM staff for their support during the campaigns. The NIKA dilution cryostat has been designed and built at the Institut N\'eel. We acknowledge the crucial contribution of the Cryogenics Group, and in particular Gregory Garde, Henri Rodenas, Jean Paul Leggeri, Philippe Camus. This work has been partially funded by the Foundation Nanoscience Grenoble and the LabEx FOCUS ANR-11-LABX-0013. This work is supported by the French National Research Agency under the contracts "MKIDS", "NIKA" and ANR-15-CE31-0017 and in the framework of the "Investissements d’avenir” program (ANR-15-IDEX-02). This work has benefited from the support of the European Research Council Advanced Grant ORISTARS under the European Union's Seventh Framework Programme (Grant Agreement no. 291294). We acknowledge fundings from the ENIGMASS French LabEx (R. A. and F. R.), the CNES post-doctoral fellowship program (R. A.), the CNES doctoral fellowship program (A. R.) and the FOCUS French LabEx doctoral fellowship program (A. R.). R.A. acknowledges support from Spanish Ministerio de Econom\'ia and Competitividad (MINECO) through grant number AYA2015-66211-C2-2. \par
}}
\vspace*{-0.3cm}

%For one-column wide figures use syntax of figure~\ref{fig-1}
%\begin{figure}[h]
%\begin{center}
%\includegraphics[scale=0.8]{myfigure.eps}
%\label{fig-1}
%\end{center}
%\end{figure}

%For two-column wide figures use syntax of figure~\ref{fig-2}
%\begin{figure*}
%\centering
%\vspace*{5cm}       % Give the correct figure height in cm
%\label{fig-2}
%\end{figure*}

%For figure with sidecaption legend use syntax of figure
%\begin{figure}
%\centering
%\sidecaption
%\includegraphics[width=5cm,clip]{tiger}
%\label{fig-3}       % Give a unique label
%\end{figure}

%For tables use syntax in table~\ref{tab-1}.
%\begin{table}
%\centering
%\label{tab-1}
% For LaTeX tables you can use
%\begin{tabular}{lll}
%\hline
%first & second & third  \\\hline
%number & number & number \\\hline
%\end{tabular}
% Or use
%\vspace*{5cm}  % with the correct table height
%\end{table}
%
% BibTeX or Biber users please use (the style is already called in the class, ensure that the "woc.bst" style is in your local directory)
% \bibliography{name or your bibliography database}

\end{document}